\documentclass[%
 aip,
 jmp,%
 amsmath,amssymb,
 reprint,%
]{revtex4-2}

\usepackage{lineno}

\usepackage{graphicx}
\usepackage{amsmath, amssymb}

\usepackage{tabularx}
\usepackage{booktabs}

\usepackage{dcolumn}
\usepackage{bm}
\usepackage[dvipsnames]{xcolor}

\usepackage{hyperref}

\hypersetup{
	breaklinks=true,
	colorlinks=true,
	citecolor=purple,
	linkcolor=purple,
	urlcolor=black,
}

\begin{document}

\preprint{AIP/123-QED}

\title{Accurate evaluation of self-heterodyne laser linewidth measurements using  Wiener filters}

\author{Markus Kantner}
\email{kantner@wias-berlin.de}
\affiliation{
Weierstrass Institute for Applied Analysis and Stochastics (WIAS),\\Mohrenstr. 39, 10117 Berlin, Germany
}%

\author{Lutz Mertenskötter}
\email{mertenskoetter@wias-berlin.de}
\affiliation{
Weierstrass Institute for Applied Analysis and Stochastics (WIAS),\\Mohrenstr. 39, 10117 Berlin, Germany
}

\date{\today}

\begin{abstract}
Self-heterodyne beat note measurements are widely used for the experimental characterization of the frequency noise power spectral density (FN--PSD) and the
spectral linewidth of lasers.
The measured data, however, must be corrected for the transfer function of the experimental setup in a post-processing routine.
The standard approach disregards the detector noise and thereby induces reconstruction artifacts, \emph{i.e.}, spurious spikes, in the reconstructed FN--PSD.
We introduce an improved post-processing routine based on a parametric Wiener filter that is free from reconstruction artifacts, provided a good estimate of the signal-to-noise ratio is supplied.
Building on this potentially exact reconstruction, we develop a new method for intrinsic laser linewidth estimation that is aimed at deliberate suppression of unphysical reconstruction artifacts.
Our method yields excellent results even in the presence of strong detector noise, where the intrinsic linewidth plateau is not even visible using the standard method.
The approach is demonstrated for simulated time series from a stochastic laser model including $1/f$-type noise.
\end{abstract}

\keywords{narrow-linewidth lasers, laser noise, colored noise, Langevin equations}
\maketitle





\section{Introduction}

Narrow-linewidth lasers exhibiting low phase noise
are core elements of coherent optical communication systems \cite{Kikuchi2016, Zhou2017, Guan2018},
gravitational wave interferometers \cite{Willke2008, Abbott2009, Dahl2019, Kapasi2020} and
emerging quantum technologies, including optical atomic clocks \cite{Camparo2007, Ludlow2015, Newman2021},
matter-wave interferometers \cite{Peters2001, Cheinet2008, Carraz2009} and ion-trap
quantum-computers \cite{Akerman2015, Bruzewicz2019, Pogorelov2021}.
For many of these applications, the performance depends critically
on the laser's intrinsic (Lorentzian) linewidth \cite{Henry1986, Wenzel2021},
which is typically obscured by additional $1/f$-like noise
\cite{Kikuchi1985,Kikuchi1989,Mercer1991,Salvade2000,Stephan2005,Spiessberger2011}. Because
of this so-called \emph{flicker noise}, the laser linewidth alone
is not a well-defined quantity and needs to be specified for a given measurement time.
For a detailed characterization of the
phase noise exhibited by the laser, the measurement of the corresponding power spectral density (PSD)
is required.

The experimental measurement of the frequency noise power
spectral density (FN--PSD) is challenging
as the rapid oscillations of the laser's optical field cannot be
 directly resolved by conventional photodetectors. A standard method
that is widely used for the characterization of the FN--PSD is the delayed self-heterodyne (DSH) beat
note technique \cite{Okoshi1980,Kikuchi1985,Dawson1992,Horak2006,Tsuchida2011,Schiemangk2014, Bai2021},
which allows to extract the phase fluctuation dynamics from a slow
beat note signal in the radio frequency (RF) regime. The  method, however, requires some post-processing of the measured
data in order to reconstruct the FN--PSD of the laser
by removing the footprint of the interferometer. 
In this paper we describe an improved post-processing routine based
on a parametric Wiener
filter that avoids typical reconstruction artifacts which occur in the standard approach. 

This paper is organized as follows: In Sec.~\ref{sec: DSH Measurement}, we describe the experimental setup and provide a model of the measurement
that takes detector noise into account. In Sec.~\ref{sec: Parametric Wiener Filter},
we review the Wiener deconvolution method with particular
emphasis on its application to DSH measurement. We discuss
a family of frequency-domain filter functions and their capabilities
in restoring the FN--PSD of the laser. In Sec.~\ref{sec: estimation method},
we present a novel method, that allows for a precise estimate of the intrinsic
linewidth even at low signal-to-noise ratio (SNR), when the onset of the
intrinsic linewidth plateau is overshadowed by measurement noise. 
The approach is demonstrated for simulated time series in Sec.~\ref{sec: simulated laser dynamics}.
We close with a discussion of the method in Sec.~\ref{sec: Discussion}.

\section{Delayed Self-Heterodyne Beat Note Measurement \label{sec: DSH Measurement}}

\begin{figure*}
\centering
\includegraphics[scale=1.1]{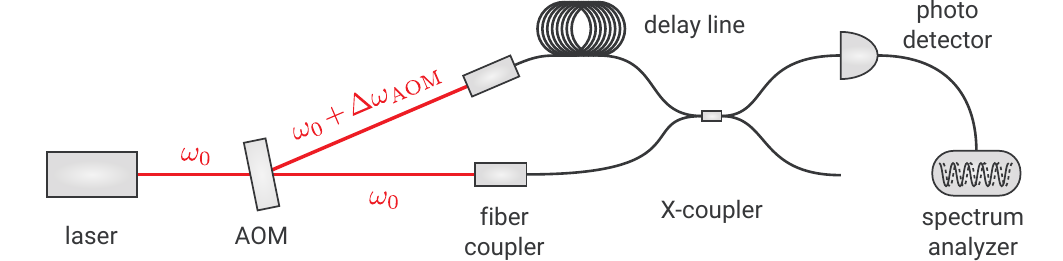}
\caption{Experimental setup for the DSH beat note measurement.
The laser beam is separated by an AOM, where one arm of the signal
is frequency shifted and delayed by a long fiber. Both beams are superimposed
at a photodetector, which captures only the slow beat note signal.}
\label{fig: DSH setup}
\end{figure*}

In the DSH measurement method, see Fig.~\ref{fig: DSH setup},
the light of a laser is superimposed with the frequency-shifted (heterodyne)
and time-delayed light from the same source. The frequency
shift $\Delta\omega_{\mathrm{AOM}}$ (typically several tens of MHz) is realized with an acousto-optic modulator (AOM)
and the delay $\tau_{d}$ is implemented via long fibers (typically
several km). If the delay is larger than the coherence time of the
laser, the delayed light can be regarded as a statistically independent
second laser with the same frequency and noise characteristics.
The DSH method allows to down-convert the optical signal to a beat note signal
in the RF domain, that can be resolved by corresponding spectrum analyzers.
Unlike other methods, the DSH method does not require stabilization
of the laser to an optical reference (\emph{e.g.}, a frequency-stabilized
second laser). 
Moreover, the frequency noise characteristics can be measured over
a broad frequency bandwidth. A detailed description of the experimental setup
and the post-processing procedure can be found in Ref.~\onlinecite{Schiemangk2019}.

After down-conversion and $I$--$Q$ demodulation (Hilbert transform) is carried
out by the spectrum analyzer, the detected in-phase and quadrature
signals read \cite{Schiemangk2019}
\begin{subequations}\label{eq: measurement IQ}
\begin{align}
I\left(t\right) & =\eta_{\mathrm{det}}\sqrt{P\left(t\right)P\left(t-\tau_{d}\right)}\,\cos{\left(\phi\left(t\right)-\phi\left(t-\tau_{d}\right)-\Delta\Omega\,t\right)}+\xi_{I}\left(t\right),\label{eq: measurement I}\\
Q\left(t\right) & =\eta_{\mathrm{det}}\sqrt{P\left(t\right)P\left(t-\tau_{d}\right)}\,\sin{\left(\phi\left(t\right)-\phi\left(t-\tau_{d}\right)-\Delta\Omega\,t\right)}+\xi_{Q}\left(t\right),\label{eq: measurement Q}
\end{align}
\end{subequations}
where $\eta_{\mathrm{det}}$ is the detector
efficiency, $P$ is the photon number, $\phi$
is the optical phase and $\Delta\Omega$ is the final difference frequency accumulated in the beating of the signal in the interferometer and the RF analyzer, where the sum frequency components are filtered out. We assume Gaussian white measurement noise with correlation function $\left\langle \xi_{I}\left(t\right)\xi_{I}\left(t'\right)\right\rangle =\left\langle \xi_{Q}\left(t\right)\xi_{Q}\left(t'\right)\right\rangle =\sigma_{\mathrm{meas}}^2\delta\left(t-t'\right)$.

From the measured time series $I\left(t\right)$, $Q\left(t\right)$
one easily obtains the phase fluctuation difference
\begin{equation}
\Delta\phi\left(t\right) = \delta\phi\left(t\right)-\delta\phi\left(t-\tau_{d}\right) =\arctan\left(\frac{Q\left(t\right)}{I\left(t\right)}\right)-\overline{\Omega}\tau_{d}+\Delta\Omega\,t+\xi_{\phi}\left(t\right)\label{eq: Delta phi time series}
\end{equation}
where $\overline{\Omega}$ is the nominal CW frequency and $\delta \phi(t) =  \phi(t) - \overline{\Omega}t$. The effective measurement noise $\xi_{\phi}\left(t\right)$ (which derives from $\xi_{I}\left(t\right)$ and $\xi_{Q}\left(t\right)$) is approximately white
\begin{equation}
\left\langle \xi_{\phi}\left(t\right)\xi_{\phi}\left(t'\right)\right\rangle \approx \left(\frac{\sigma_{\mathrm{meas}}}{\eta_{\mathrm{det}}\overline{P}}\right)^{2}\delta\left(t-t'\right),\label{eq: phase measurement noise}
\end{equation}
if the average power $\overline{P}$ is much larger than the measurement noise level $\sigma_{\mathrm{meas}}$, see Appendix~\ref{sec: phase measurement noise}.
The evaluation of Eq.~\eqref{eq: Delta phi time series} requires estimation
of $\tau_{d}$ and $\Delta\Omega$ (detrending), see \cite{Schiemangk2019} for details.

In the frequency domain, the relation between the phase fluctuations  $\delta\phi\left(t\right)$ and
$\Delta\phi\left(t\right)$ reads
\begin{equation}\Delta\tilde{\phi}\left(\omega\right)=H\left(\omega\right)\delta\tilde{\phi}\left(\omega\right),\qquad H\left(\omega\right)=1-\mathrm{e}^{i\omega\tau_{d}},\label{eq: transfer function}
\end{equation}
from which one derives a simple relation between the corresponding phase noise PSDs
\begin{equation}
S_{\Delta\phi,\Delta\phi}\left(\omega\right)=\left|H\left(\omega\right)\right|^{2}S_{\delta\phi,\delta\phi}\left(\omega\right).\label{eq: relation between power spectra}
\end{equation}
In the standard post-processing routine \cite{Kikuchi1985, Schiemangk2019}, Eq.~\eqref{eq: relation between power spectra}
is solved for $S_{\delta\phi,\delta\phi}\left(\omega\right)$ by division through
$\left|H\left(\omega\right)\right|^{2}=2\left(1-\cos{\left(\omega\tau_{d}\right)}\right)$.
This approach has two notable shortcomings: First, this procedure
does not take into account the detector noise
and thereby fails at increased measurement
noise levels. Second, the transfer function has roots at $\omega_{n}=2\pi n/\tau_{d}$,
$n\in\mathbb{Z}$, which turn to poles in its inverse $\left|G\left(\omega\right)\right|^{2}=\left|H\left(\omega\right)\right|^{-2}$.
Hence, the reconstructed PSD exhibits a series of
equidistant spurious spikes \cite{LewoczkoAdamczyk2015,Wenzel2022, Kumar2022}, resulting
from an uncontrolled amplification of the measurement noise.

\section{Parametric Wiener Filters\label{sec: Parametric Wiener Filter}}

In this section, we present the Wiener deconvolution method for reconstructing hidden signals from noisy time series data. Besides the well-known Wiener filter, we introduce \emph{power spectrum equalization} (PSE) as an important representative of the group of parametric Wiener filters \cite{Lim1990}.

Let $x\left(t\right)$ denote the time series of a hidden signal of
interest that is measured by an experimental setup characterized by
a convolution kernel $h\left(t\right)$. In the case of the DSH measurement described above, this is $h\left(t\right)=\delta\left(t\right)-\delta\left(t-\tau_{d}\right)$.
Furthermore, let $\xi\left(t\right)$ denote additive Gaussian white
measurement noise. Then the experiment yields an observed time series
\begin{subequations}\label{eq: Wiener deconvolution}
\begin{equation}
z\left(t\right)=\left(h*x\right)\left(t\right)+\xi\left(t\right).\label{eq: Wiener deconvolution z}
\end{equation}
The process noise and measurement noise are assumed to be
uncorrelated $\langle x\left(t\right)\xi\left(t'\right)\rangle=0$.
One seeks for an optimal estimate $\hat{x}\left(t\right)$ of the hidden signal
\begin{equation}
\hat{x}\left(t\right)=\left(g*z\right)\left(t\right), \label{eq: Wiener deconvolution x estimate}
\end{equation}
\end{subequations}
where the (de-)convolution kernel $g\left(t\right)$ minimizes the reconstruction error.

In this paper, our main interest is the reconstruction of PSDs of
hidden signals in the frequency domain, for which we introduce the
Fourier space representation of Eq.~\eqref{eq: Wiener deconvolution}
\begin{subequations}\label{eq: Wiener deconvolution - Fourier space}
\begin{align}
Z\left(\omega\right) & =H\left(\omega\right)X\left(\omega\right)+\Xi\left(\omega\right),\label{eq: Z Fourier}\\
\hat{X}\left(\omega\right) & =G\left(\omega\right)Z\left(\omega\right).\label{eq: Xhat Fourier}
\end{align}
\end{subequations}
From Eq.~\eqref{eq: Xhat Fourier}, we obtain the relation
between the estimated PSD $S_{\hat{x},\hat{x}}\left(\omega\right)$
of the hidden signal and the PSD of the measured time series $S_{z,z}\left(\omega\right)$
\begin{equation}
S_{\hat{x},\hat{x}}\left(\omega\right)=\left|G\left(\omega\right)\right|^{2}S_{z,z}\left(\omega\right). \label{eq: estimated power spectrum S_xhat_xhat}
\end{equation}

\begin{figure}[t]
\centering
\includegraphics[scale=1.1]{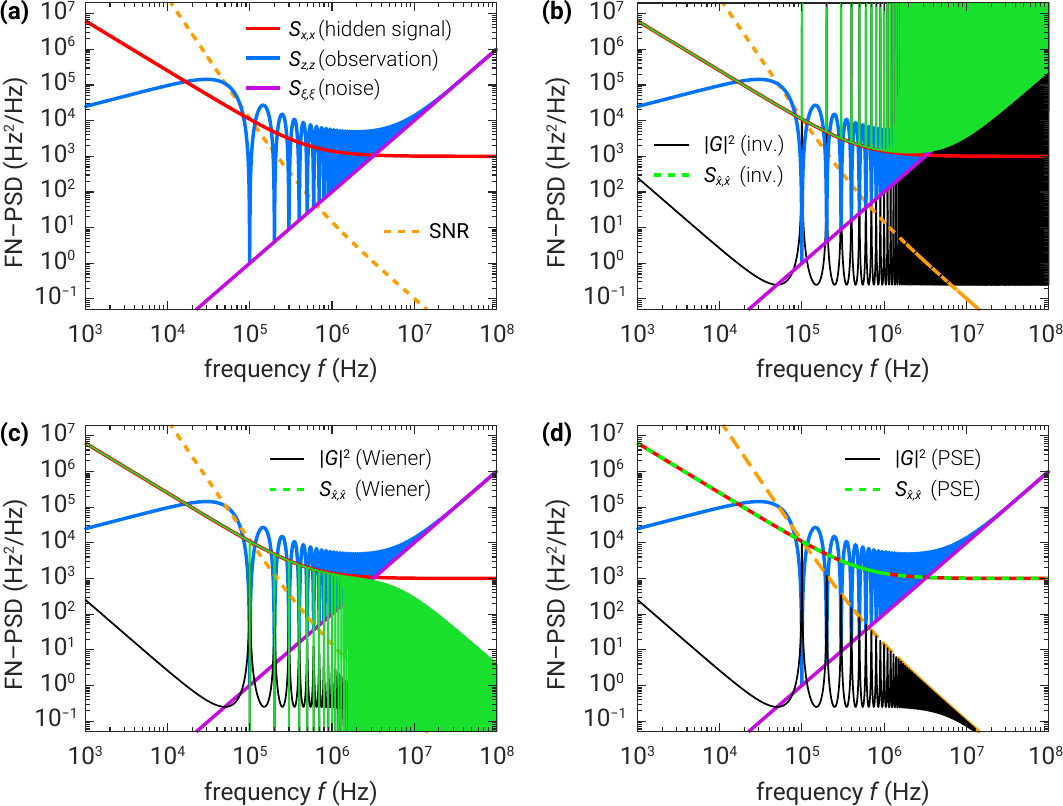}
\caption{Comparison of different filters $G\left(\omega\right)$ for FN--PSD reconstruction.
\textbf{(a)}~Analytical signal Eq.~\eqref{eq: analytical signal S_xx} and
measurement noise PSD \eqref{eq: analytical signal S_xixi} along with the observed
spectrum and the exact SNR. Parameters in the plot
are $\nu=1.4$, $C=10^{11}\,\mathrm{Hz}^{\nu+1}$, $S_{\infty}=10^{3}\,\mathrm{Hz}$,
$\sigma=10^{-10}\,\mathrm{Hz}$ and $\tau_{d}=10$\,\textmu$\mathrm{s}$.
\textbf{(b)}~The inverse filter Eq.~\eqref{eq: G inverse} yields a reconstructed
PSD with spurious peaks at the pole frequencies $f_{n}=n/\tau_{d}$,
$n\in\mathbb{Z}$. The intrinsic linewidth
plateau is obscured by detector noise and can not be recovered from the inverse
filter (merely an upper limit can be extracted).
\textbf{(c)}~The Wiener filter Eq.~\eqref{eq: standard Wiener filter} is
optimized for
time series reconstruction, but fails in reconstruction of the PSD.
The reconstructed PSD features sharp dropouts at the pole frequencies
and does not follow the hidden signal at low $\mathrm{SNR}<1$. \textbf{(d)}~Power
spectrum equalization Eq.~\eqref{eq: power spectrum equalization} yields
an exact reconstruction of the hidden signal's PSD if the exact
SNR is provided. Here, the singularities at the pole frequencies are bounded from above by the SNR, which allows for
exact compensation of both the detector noise and the interferometer effects in the observed signal $S_{z,z}$.}
\label{fig: filter comparison}
\end{figure}

In the following, we discuss different candidates for the 
filter function $G\left(\omega\right)$.
Their performance is assessed
with regard to the reconstruction of the FN--PSD of a semiconductor
laser \cite{Kikuchi1989,Salvade2000,Stephan2005} from DSH measurements.
The transfer function of the interferometer is
\begin{equation*}
H\left(\omega\right)=1-\mathrm{e}^{i\omega\tau_{d}}
\end{equation*}
and we assume the hidden signal and noise PSDs as
\begin{align}
S_{x,x}\left(\omega\right) & =\frac{C}{\omega^{\nu}}+S_{\infty},\label{eq: analytical signal S_xx}\\
S_{\xi,\xi}\left(\omega\right) & =\sigma\omega^{2}.\label{eq: analytical signal S_xixi}
\end{align}
In Eq.~\eqref{eq: analytical signal S_xx}, $S_{\infty}$ determines the
intrinsic laser linewidth, which is obscured by additional
colored noise of power-law type with $0.8\lesssim\nu\lesssim 1.6$ (flicker
noise). The functional form of Eq.~\eqref{eq: analytical signal S_xx} is
consistent with theoretical models and experimental observations for
frequencies well below the relaxation oscillation (RO) peak (typically at
several GHz). The level of phase measurement noise, cf. Eq.~\eqref{eq: phase measurement noise}, is specified by $\sigma$ and the corresponding frequency
measurement noise PSD is a quadratic function of the frequency.
The model PSDs \eqref{eq: analytical signal S_xx}--\eqref{eq: analytical signal S_xixi}
imply the signal-to-noise ratio 
\begin{equation}
\mathrm{SNR}\left(\omega\right)=S_{x,x}\left(\omega\right)/S_{\xi,\xi}\left(\omega\right).\label{eq: SNR}
\end{equation}
Figure~\ref{fig: filter comparison} shows that different filters
$G\left(\omega\right)$ can lead to vastly different results for $S_{\hat{x},\hat{x}}\left(\omega\right)$. In the following section, we discuss their behavior in more detail.

\subsection{Inverse Filter}

In the case of negligible detector noise, the filter $G\left(\omega\right)$
is given by the inverse transfer function
\begin{equation}
G_{\mathrm{inv}}\left(\omega\right)=H^{-1}\left(\omega\right).\label{eq: G inverse}
\end{equation}
The corresponding estimate of the PSD of the hidden signal reads
\begin{align*}
S_{\hat{x},\hat{x}}\left(\omega\right) & =\left|G_{\mathrm{inv}}\left(\omega\right)\right|^{2}S_{z,z}\left(\omega\right), & \left|G_{\mathrm{inv}}\left(\omega\right)\right|^{2} & = \left|H\left(\omega\right)\right|^{-2},
\end{align*}
which coincides with the standard post-processing method 
of the DSH measurement \cite{Kikuchi1985, Schiemangk2014, Wenzel2022}, cf. Eq.~\eqref{eq: relation between power spectra}. 
The most prominent feature of the inverse filter $\left|G_{\mathrm{inv}}\left(\omega\right)\right|^{2}$
are singularities at the poles $\omega_{n}^{\mathrm{pole}}=2\pi n/\tau_{d}$,
$n\in\mathbb{Z}$, where the PSD reconstruction fails, see Fig.~\ref{fig: filter comparison}\,(b).
Sufficiently far away from these poles, the reconstructed spectrum matches
the hidden signal as long as the signal-to-noise ratio is large ($\mathrm{SNR}>1$).
If the intrinsic linewidth plateau is obscured by measurement noise,
only an upper limit can be extracted via inverse filtering.

\subsection{Wiener Filter}

Wiener filtering achieves an optimal trade-off between inverse filtering
and noise removal. It subtracts the additive noise and reverses the
effects of the interferometer simultaneously. The Wiener filter is
obtained from minimizing the mean square error of the time-domain
signal at an arbitrary instance of time, see Appendix~\ref{sec: Wiener Filter derivation}.
In the frequency domain, the Wiener filter reads
\begin{align}
\begin{aligned}
G_{\mathrm{Wiener}}\left(\omega\right)  &=\frac{H^{*}\left(\omega\right)S_{x,x}\left(\omega\right)}{\left|H\left(\omega\right)\right|^{2}S_{x,x}\left(\omega\right)+S_{\xi,\xi}\left(\omega\right)}  \\
&=\frac{1}{H\left(\omega\right)}\left(1+\frac{1}{\left|H\left(\omega\right)\right|^{2}\mathrm{SNR}\left(\omega\right)}\right)^{-1}.
\end{aligned}\label{eq: standard Wiener filter}
\end{align}
Although the Wiener filter provides an optimal reconstruction of the
time-domain signal, the corresponding PSD reconstruction deviates
significantly from the true spectrum in regions of low SNR, see Fig.~\ref{fig: filter comparison}\,(c).
Moreover, we note that the Wiener filter overemphasizes noise reduction
at the poles $\omega_{n}^{\mathrm{pole}}$, where the reconstructed
PSD is zero because of 
$\big|G_{\mathrm{Wiener}}\big(\omega_{n}^{\mathrm{pole}}\big)\big|^{2}=0$,
such that also $S_{\hat{x},\hat{x}}\big(\omega_{n}^{\mathrm{pole}}\big)=0$.
Away from these poles and at high SNR, the Wiener filter asymptotically
approaches the behavior of the inverse filter:
$\big|G_{\mathrm{Wiener}}\big(\omega\neq\omega_{n}^{\mathrm{pole}}\big)\big|^{2}\stackrel{\mathrm{SNR}\to\infty}{\sim} \left|H\left(\omega\right)\right|^{-2}$.

\subsection{Power Spectrum Equalization}

Besides the standard Wiener filter, there exist several variants
of the method which are collectively referred to as parametric Wiener
filters \cite{Lim1990}. An important one is \emph{power spectrum
equalization} (PSE), which is tailored to minimize the quadratic error
of the reconstructed PSD, see Appendix~\ref{sec: PSE Filter derivation}.
The corresponding filter function reads
\begin{align}
\begin{aligned}
\bigl|G_{\mathrm{PSE}}\left(\omega\right)\bigr|^{2}  &= \frac{S_{x,x}\left(\omega\right)}{\left|H\left(\omega\right)\right|^{2}S_{x,x}\left(\omega\right)+S_{\xi,\xi}\left(\omega\right)} \\
&=\frac{1}{\left|H\left(\omega\right)\right|^{2}}\left(1+\frac{1}{\left|H\left(\omega\right)\right|^{2}\mathrm{SNR}\left(\omega\right)}\right)^{-1}.
\end{aligned}\label{eq: power spectrum equalization}
\end{align}
The PSE filter yields an accurate reconstruction of the hidden
signal when the true frequency-dependent SNR is provided, see Fig.~\ref{fig: filter comparison}\,(d).

Most remarkably, the reconstructed spectrum is free of reconstruction artifacts at
the poles of the inverse filter function. This result is easily understood
by the following analysis. A straightforward calculation shows that
the filter approaches the SNR at $\omega_{n}^{\mathrm{pole}}$:
$\big|G_{\mathrm{PSE}}\big(\omega_{n}^{\mathrm{pole}}\big)\big|^{2}  =\mathrm{SNR}\big(\omega_{n}^{\mathrm{pole}}\big)$.
As the interferometer is blind for these frequency components (\emph{i.e.},
the transfer function is zero $H\big(\omega_{n}^{\mathrm{pole}}\big)=0$),
the observed signal contains only measurement noise $S_{z,z}\big(\omega_{n}^{\mathrm{pole}}\big)=S_{\xi,\xi}\big(\omega_{n}^{\mathrm{pole}}\big)$, see Eq.~\eqref{eq: Xhat Fourier}. Finally, substitution into
Eq.~\eqref{eq: estimated power spectrum S_xhat_xhat}, shows that
the PSE filter cancels out the measurement noise exactly and recovers
the true signal
\begin{equation*}
S_{\hat{x},\hat{x}}\big(\omega_{n}^{\mathrm{pole}}\big) =\bigl|G_{\mathrm{PSE}}\big(\omega_{n}^{\mathrm{pole}}\big)\bigr|^{2}\,S_{z,z}\big(\omega_{n}^{\mathrm{pole}}\big)  =\mathrm{SNR}\big(\omega_{n}^{\mathrm{pole}}\big)\,S_{\xi,\xi}\big(\omega_{n}^{\mathrm{pole}}\big)=S_{x,x}\big(\omega_{n}^{\mathrm{pole}}\big)
\end{equation*}
if the correct SNR is provided. Furthermore, we observe that the PSE
filter restores the hidden signal even in regions of low SNR. This
result follows along the same lines as above, starting from $\bigl|G_{\mathrm{PSE}}\left(\omega\right)\bigr|^{2}\stackrel{\mathrm{SNR}\to 0}{\sim}\mathrm{SNR}\left(\omega\right)$.
In the opposite case, at very high $\mathrm{SNR}\left(\omega\right) \gg 1$,
the PSE filter again approaches (just like the Wiener filter) the
inverse filter $\bigl|G_{\mathrm{PSE}}\left(\omega\right)\bigr|^{2}\stackrel{\mathrm{SNR} \to\infty}{\sim}\left|H\left(\omega\right)\right|^{-2}.$

\vspace{1em}

Finally, we note that all the filter candidates discussed in this
section can be written in a unified way as parametric Wiener filters
of the following form:
\begin{equation*}
\bigl|G\left(\omega\right)\bigr|^{2}=\frac{1}{\left|H\left(\omega\right)\right|^{2}}\left(1+\frac{1}{\left|H\left(\omega\right)\right|^{2}\mathrm{SNR}\left(\omega\right)}\right)^{-m}=\begin{cases}
\bigl|G_{\mathrm{inv}}\left(\omega\right)\bigr|^{2} & \text{for }m=0,\\
\bigl|G_{\mathrm{Wiener}}\left(\omega\right)\bigr|^{2} & \text{for }m=1,\\
\bigl|G_{\mathrm{PSE}}\left(\omega\right)\bigr|^{2} & \text{for }m=2.
\end{cases}
\end{equation*}

\section{Intrinsic Linewidth Estimation at Low Signal-to-Noise
Ratio\label{sec: estimation method}}

\begin{figure}
\centering
\includegraphics[scale=1.1]{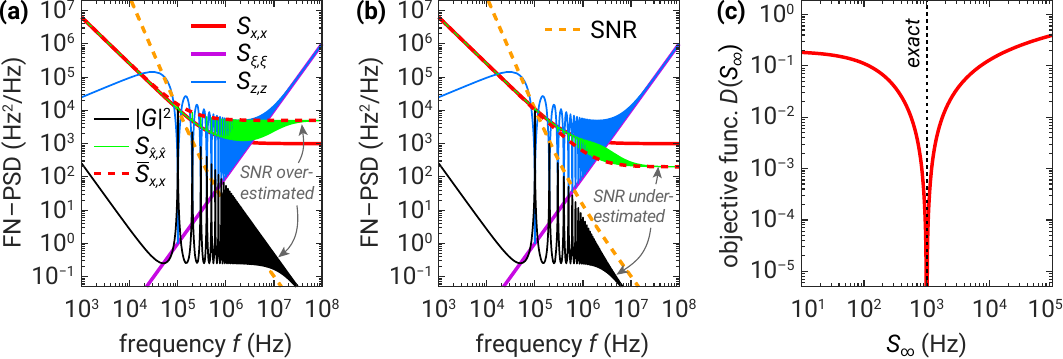}
\caption{Reconstruction artifacts in the PSE filter method with
incorrectly estimated SNR.
\textbf{(a)}~Overestimation ($S_{\infty}^{\mathrm{est}}=5S_{\infty}$)
and \textbf{(b)} underestimation ($S_{\infty}^{\mathrm{est}}=0.2S_{\infty}$)
of the intrinsic linewidth $S_\infty$ leads to spurious oscillations and spikes
in the reconstructed
spectrum $S_{\hat{x},\hat{x}}$. In the 
case of misspecification of the SNR, the maxima of 
$\left|G_{\mathrm{PSE}}\right|^{2}$ are no longer bounded by the exact SNR. The assumed $\overline{S}_{x,x}$, which has the functional
form \eqref{eq: analytical signal S_xx} and enters the SNR estimate,
is shown as a red dashed line. 
The method described in Sec.~\ref{sec: estimation method}
aims at minimizing the deviation between $\overline{S}_{x,x}$ and
$S_{\hat{x},\hat{x}}$ in order to estimate the true value of the
intrinsic linewidth parameter $S_{\infty}$.  \textbf{(c)}~The corresponding objective function \eqref{eq: objective function} features a sharp minimum at the exact value.}
\label{fig: PSE over and under estimation}
\end{figure}

In the previous section, it was shown that the PSE filter can provide
an excellent reconstruction of the hidden signal's PSD if the exact SNR is supplied.
At first glance, this approach appears to be rather impractical,
since the specification of the exact SNR already anticipates the actual
measurement result to a certain degree.
One might therefore worry that arbitrary reconstructions could be generated.
It turns out, however, that the PSE filter method introduces 
characteristic reconstruction artifacts when the specified SNR is incorrect, 
see Fig.~\ref{fig: PSE over and under estimation}. These
spurious spikes are easily recognized to be unphysical, such that
the incorrect SNR estimate can be rejected.
Based on this observation,  we develop a method that simultaneously reconstructs
both the PSD of the hidden signal as well as the correct SNR,
by minimizing these reconstruction artifacts.

In the following, we employ again the analytic model PSDs \eqref{eq: analytical signal S_xx}--\eqref{eq: analytical signal S_xixi}.
For the sake of simplicity,
we assume that the parameters $C$ and $\nu$ can
be accurately estimated from the data, since the low-frequency part
of the signal is only negligibly affected by measurement noise.
Similarly, we assume that the noise level $\sigma$ is known from
independent noise floor measurements or from analysis of the relative intensity noise (RIN) PSD, which is typically dominated by measurement noise at increased powers.
The only free parameter to be estimated then is $S_{\infty}$.

Figure~\ref{fig: PSE over and under estimation}\,(a)--(b) shows
the effects of over- and underestimation of $S_{\infty}$ in the analytical
model. Due to the mismatch between the filter function $\left|G_{\mathrm{PSE}}\left(\omega\right)\right|^{2}$
and the observed spectrum $S_{z,z}\left(\omega\right)$, spurious
spikes (reconstruction artifacts) show up at frequencies $\omega\approx\omega_{n}^{\mathrm{pole}}$
in the reconstructed spectrum $S_{\hat{x},\hat{x}}\left(\omega\right)$.
At large frequencies these spikes are damped out in both $\left|G_{\mathrm{PSE}}\left(\omega\right)\right|^{2}$ and
$S_{z,z}\left(\omega\right)$, but their product yields a wrong value
of the intrinsic linewidth plateau. We introduce an objective function $D\left(S_{\infty}\right)$
that penalizes this deviation (\emph{i.e.}, the ``inconsistency'')
between the reconstructed signal $S_{\hat{x},\hat{x}}\left(\omega;S_{\infty}\right)$
(depending on the assumed SNR as a function of estimated $S_{\infty}$)
and the implicitly assumed signal $\overline{S}_{x,x}\left(\omega;S_{\infty}\right)$
obeying the functional form \eqref{eq: analytical signal S_xx} as
\begin{equation}
D\left(S_{\infty}\right)=\left(\int\mathrm{d}\omega\,\frac{S_{\hat{x},\hat{x}}\left(\omega;S_{\infty}\right)-\overline{S}_{x,x}\left(\omega;S_{\infty}\right)}{\overline{S}_{x,x}\left(\omega;S_{\infty}\right)}\right)^{2} \label{eq: objective function}
\end{equation}
where $S_{\hat{x},\hat{x}}\left(\omega;S_{\infty}\right)=\left|G_{\mathrm{PSE}}\left(\omega;S_{\infty}\right)\right|^{2}S_{z,z}\left(\omega\right)$.
The $\omega$--integral runs over a suitable frequency range.
As shown in Fig.~\ref{fig: PSE over and under estimation}\,(c), the objective function \eqref{eq: objective function}
exhibits a sharp minimum at the exact value, cf.~Fig.~\ref{fig: filter comparison}\,(c).
Hence, $S_\infty$ can be estimated by minimization of $D(S_\infty)$.

\section{Application to Stochastic Laser Dynamics\label{sec: simulated laser dynamics}}

In this section, we demonstrate the method described in Sec.~\ref{sec: estimation method} for simulated time series.
In Sec.~\ref{sec: Stochastic Laser Rate Equations}, we introduce a stochastic laser model including non-Markovian colored noise, that generates realistic time series with frequency drifts as commonly observed for diode lasers.
In Sec.~\ref{sec: application to simulated data}, we apply the linewidth estimation method to simulated DSH measurement data.

\subsection{Stochastic Laser Rate Equations\label{sec: Stochastic Laser Rate Equations}}

\begin{figure}[t]
\centering
\includegraphics[scale=1.1]{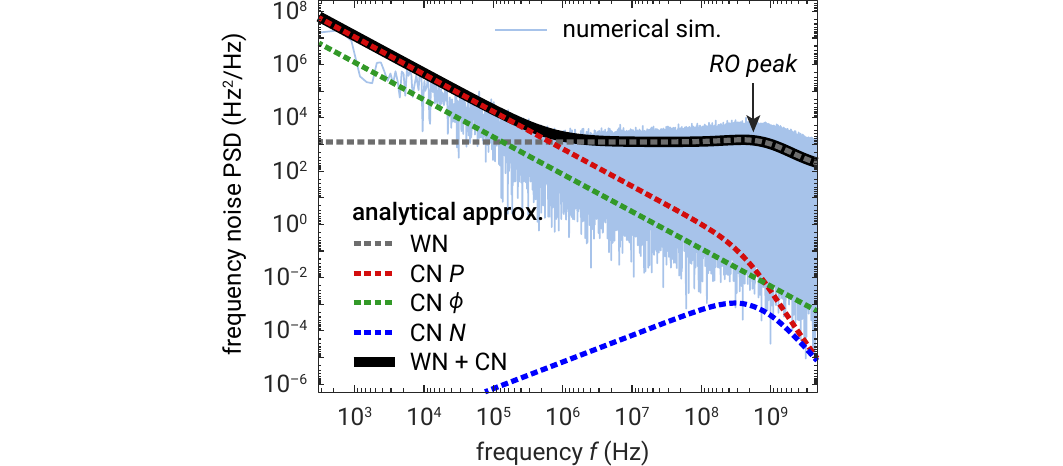}
\caption{Numerically computed FN--PSD from simulated time series using the stochastic laser rate Eqs.~\eqref{eq: Langevin laser model}. Analytical approximations are derived from linearization of the model at the noise-free steady state. White noise (WN) and different colored noise (CN) contributions are indicated separately.
}
\label{fig: numerical FN-PSD}
\end{figure}

We consider a  Langevin equation model for a generic single-mode semiconductor laser
\begin{subequations}\label{eq: Langevin laser model}
\begin{align}
\dot{P} & =-\gamma\left(P-P_{\mathrm{th}}\right)+\Gamma v_{g}g\left(P,N\right)P+\Gamma v_{g}g_{\mathrm{sp}}\left(P,N\right)+F_{P},\label{eq: langevin P}\\
\dot{\phi} & =\Omega_{0}+\frac{\alpha_{H}}{2}\Gamma v_{g}g\left(P,N\right)+F_{\phi},\label{eq: langevin phi}\\
\dot{N} & =\frac{\eta I}{q}-R\left(N\right)-\Gamma v_{g}g\left(P,N\right)P-\Gamma v_{g}g_{\mathrm{sp}}\left(P,N\right)+F_{N},\label{eq: langevin N}
\end{align}
\end{subequations}
where $P$ is the number of photons, $\phi$ is the optical phase and $N$ is the number of charge carriers in the active region.
Moreover, $\gamma$ is the inverse photon lifetime, $P_{\mathrm{th}}$
is the thermal photon number (Bose--Einstein factor), $\Gamma$ is
the optical confinement factor, $v_{g}$ is the group velocity, $\Omega_{0}$
is the detuning from the CW reference frequency, $\alpha_{H}$ is the
linewidth enhancement factor, $I$ is the pump current,
$\eta$ is the injection efficiency and $q$ is the elementary charge.
The net-gain is modeled as
\begin{equation}
g\left(P,N\right)=\frac{g_{0}}{1+\varepsilon P}\log{\left(\frac{N}{N_{\mathrm{tr}}}\right)},\label{eq: net-gain}
\end{equation}
where $g_{0}$ is the gain coefficient, $N_{\mathrm{tr}}$ is the carrier
number at transparency and $\varepsilon$ is the gain compression
coefficient. Following \cite{Wenzel2021}, the spontaneous
emission coefficient is described by 
\begin{equation}
g_{\mathrm{sp}}\left(P,N\right)=\frac{1}{2}\frac{g_{0}}{1+\varepsilon P}\log{\left(1+\left(\frac{N}{N_{\mathrm{tr}}}\right)^{2}\right)},\label{eq: G spont}
\end{equation}
which does not require any additional parameters and avoids the introduction of the population inversion factor \cite{Henry1986,Wenzel2021}.
The stimulated absorption coefficient
is implicitly given by Eqs.~\eqref{eq: net-gain}--\eqref{eq: G spont}
as $g_{\mathrm{abs}}\left(P,N\right)=g_{\mathrm{sp}}\left(P,N\right)-g\left(P,N\right)$.
Non-radiative recombination and spontaneous emission into waste modes
are described by
\begin{equation}
R\left(N\right)=AN+\frac{B}{V}N^{2}+\frac{C}{V^{2}}N^{3},\label{eq: recombination R}
\end{equation}
where $A$ is the Shockley--Read--Hall recombination rate, $B$
is the bimolecular recombination coefficient, $C$ is the Auger recombination
coefficient and $V$ is the volume of the active region.

\begin{table}[t]
\centering
\renewcommand{\arraystretch}{1.0}
\begin{tabular}{cll}
\toprule 
\textbf{symbol} & \textbf{description} & \textbf{value}\\
\midrule
$\gamma$ & inverse photon lifetime & $5\cdot10^{11}\,\mathrm{s}^{-1}$\\
$P_{\mathrm{th}}$ & thermal photon number & $2.7\cdot10^{-20}$\\
$\Gamma$ & optical confinement factor & $0.01$\\
$g_{0}$ & gain coefficient & $3.54\cdot10^{5}\,\mathrm{m}^{-1}$\\
$n_{g}$ & group index & $3.9$\\
$v_{g}$ & group velocity, $v_{g}=n_{g}/c_{0}$ & $7.69\cdot10^{7}\,\mathrm{m}\mathrm{s}^{-1}$\\
$N_{\mathrm{tr}}$ & transparency carrier number & $2.5\cdot10^{9}$\\
$\varepsilon$ & gain compression coefficient & $10^{-8}$\\
$\Omega_{0}$  & detuning from CW reference freq. & $0\,\mathrm{Hz}$\\
$\alpha_{H}$ & linewidth enhancement factor & $3.0$\\
$I$ & pump current & $200\,\mathrm{mA}$\\
$\eta$ & injection efficiency & $0.9$\\
$A$ & Shockley--Read--Hall recombination rate & $1\cdot10^{8}\,\mathrm{s}^{-1}$\\
$B$ & bimolecular recombination coefficient & $1\cdot10^{16}\,\mathrm{m}^{3}\mathrm{s}^{-1}$\\
$C$ & Auger recombination coefficient & $4\cdot10^{-42}\,\mathrm{m}^{6}\mathrm{s}^{-1}$\\
$V$ & active region volume & $1.25\cdot10^{-15}\,\mathrm{m}^{3}$\\
$\nu_{P}$ & colored noise exponent & $1.4$\\
$\sigma_{P,0}$ & colored noise amplitude & $5\cdot10^{5}\,\mathrm{s}^{-(1+\nu_{P})/2}$\\
$\nu_{N}$ & colored noise exponent & $1.0$\\
$\sigma_{N,0}$ & colored noise amplitude & $10^{9}\,\mathrm{s}^{-(1+\nu_{N})/2}$\\
$\sigma_{\mathrm{meas}}$ & detector noise floor level & $2\cdot10^{3}\,\mathrm{s}^{1/2}\,\eta_{\mathrm{det}}$\\
$\tau_{d}$ & interferometer delay & $10\cdot10^{-6}\,\mathrm{s}$\\
\bottomrule
\end{tabular}
\caption{List of parameter values used in stochastic time series simulation.}
\label{tab: parameter values}
\end{table}

The Langevin forces describe zero-mean Gaussian colored noise with the following non-vanishing frequency-domain correlation functions:
\begin{align}
\begin{aligned}
\langle\tilde{F}_{P}\left(\omega\right)\tilde{F}_{P}\left(\omega'\right)\rangle & =\left(2\left(\Gamma v_{g}g_{\mathrm{sp}}\big(\overline{P},\overline{N}\big)+\gamma P_{\mathrm{th}}\right)\overline{P}\left(1+\frac{1}{\overline{P}}\right)+\sigma_{P}^{2}\big(\overline{P}\big)\frac{1}{\omega^{\nu_{P}}}\right)\,\delta\left(\omega-\omega'\right),\\
\langle\tilde{F}_{\phi}\left(\omega\right)\tilde{F}_{\phi}\left(\omega'\right)\rangle & =\left(\left(\Gamma v_{g}g_{\mathrm{sp}}\big(\overline{P},\overline{N}\big)+\gamma P_{\mathrm{th}}\right)\left(1+\frac{1}{\overline{P}}\right)+\left(\frac{\sigma_{P}\big(\overline{P}\big)}{2\overline{P}}\right)^{2}\frac{1}{\omega^{\nu_{P}}}\right)\,\delta\left(\omega-\omega'\right),\\
\langle\tilde{F}_{N}\left(\omega\right)\tilde{F}_{N}\left(\omega'\right)\rangle & =\left(2R\big(\overline{N}\big)+2\Gamma v_{g}g_{\mathrm{sp}}\big(\overline{P},\overline{N}\big)\overline{P}\left(1+\frac{1}{\overline{P}}\right)+\frac{\sigma_{N}^{2}\big(\overline{N}\big)}{\omega^{\nu_{N}}}\right)\,\delta\left(\omega-\omega'\right),\\
\langle\tilde{F}_{P}\left(\omega\right)\tilde{F}_{N}\left(\omega'\right)\rangle & =-\left(\Gamma v_{g}g_{\mathrm{sp}}\big(\overline{P},\overline{N}\big)\left(2\overline{P}+1\right)-\Gamma v_{g}g\big(\overline{P},\overline{N}\big)\overline{P}\right)\,\delta\left(\omega-\omega'\right),
\end{aligned}\label{eq: Langevin correlation functions}
\end{align}
The white noise part of the model includes a quantum
mechanically consistent description of light-matter
interaction fluctuations \cite{Coldren2012}. Moreover, we have included three
independent $1/f$-type noise sources
with power-law exponents $\nu_{P}$ and $\nu_{N}$,
respectively. The colored noise amplitudes are taken as $\sigma_{P}\left(P\right)=2P\sigma_{P,0}$
and $\sigma_{N}\left(N\right)=\sqrt{N}\sigma_{N,0}$ (modeling Hooge's law \cite{Hooge1994, Garmash1989}).
The noise correlation functions~\eqref{eq: Langevin correlation functions}
are formulated at the unique noise-free steady state $\big(\overline{P},\overline{N}\big)$. The full nonlinear system of Itô-type stochastic differential
equations used for simulation is given in Appendix~\ref{sec: Ito SDEs}.
The numerically simulated FN--PSD is shown in Fig.~\ref{fig: numerical FN-PSD} along with (semi-)analytical approximations.
All parameter values used in the simulations are listed in Tab.~\ref{tab: parameter values}.

\subsection{Intrinsic Linewidth Estimation\label{sec: application to simulated data}}

\begin{figure}[t]
\centering
\includegraphics[scale=1.1]{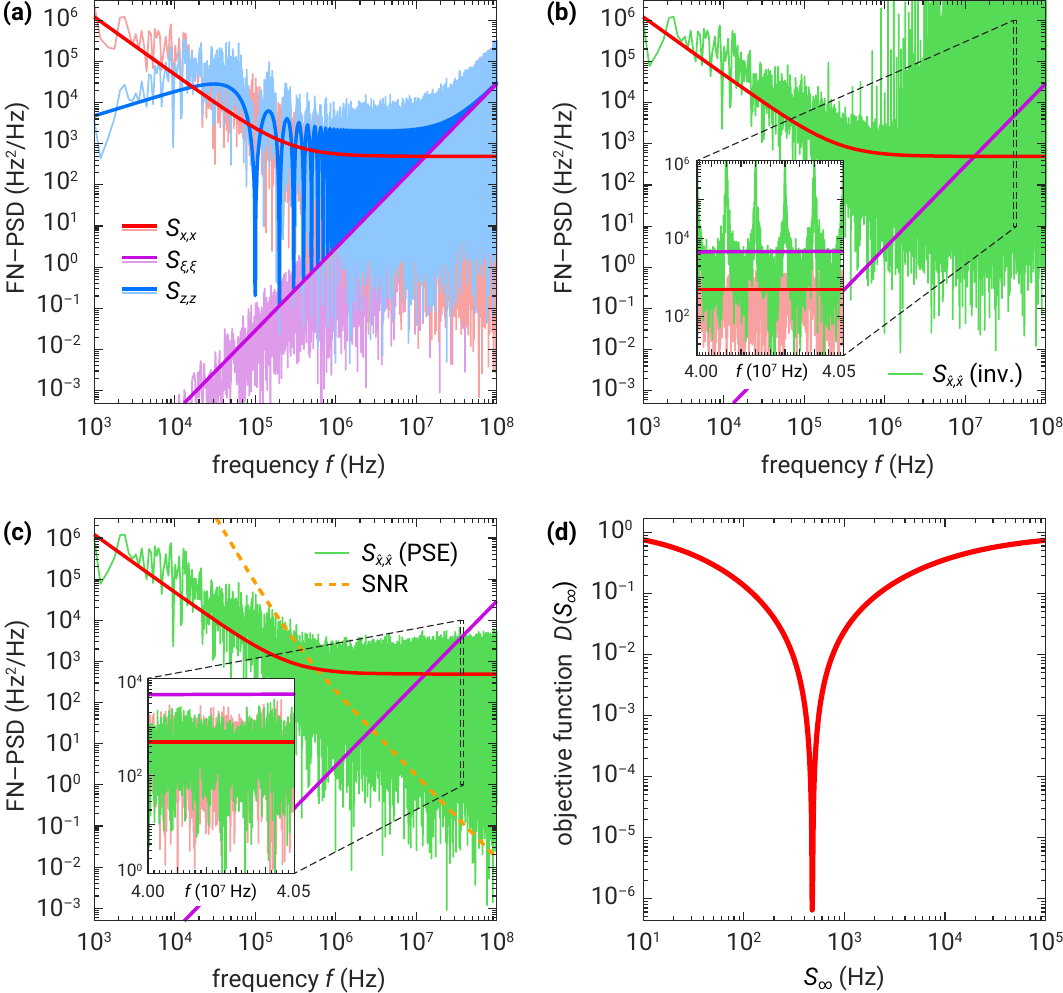}
\caption{Application of the linewidth estimation method to simulated time series data. \textbf{(a)}~PSDs of the hidden signal, the measurement noise and the measured data along with their analytic values. \textbf{(b)}~Reconstructed PSD using the inverse filter. The inset zooms in on a region with low SNR, where the reconstructed and the true signal deviate by about one order of magnitude. Moreover, we observe reconstruction artifacts at the pole frequencies. \textbf{(c)}~The PSE filter method yields an accurate reconstruction of the hidden signal even at low SNR that is free of reconstruction artifacts. Here the analytic model PSDs were fitted to the PSD to give the SNR according to the method described in Sec.~\ref{sec: estimation method}. \textbf{(d)}~Minimization of the objective function~\eqref{eq: objective function} yields a sharp estimate of the intrinsic linewidth parameter at $S_\infty \approx 480\,\mathrm{Hz}$.}
\label{fig: method applied to simulation data}
\end{figure}

We apply the method described in Sec.~\ref{sec: estimation method}
to simulated DSH measurements.
The simulation is carried out in two
steps: First, the stochastic laser rate Eqs.~\eqref{eq: Langevin laser model} are simulated using the Euler--Maruyama
method (time step $\Delta t=50\,\mathrm{ps}$).
In the second step, the
DSH measurement is simulated by evaluation of Eq.~\eqref{eq: measurement IQ},
which includes addition of Gaussian white measurement noise. The simulated $I$--$Q$
data are used to generate the time series $\Delta\phi$ according to
Eq.~\eqref{eq: Delta phi time series}. The observed spectrum is 
computed from $S_{z,z}\left(\omega\right)=\omega^{2}S_{\Delta\phi,\Delta\phi}\left(\omega\right)$ and shown in Fig.~\ref{fig: method applied to simulation data}\,(a).
For recovery of the original FN--PSD, the PSE filter method is applied
to the simulated FN--PSD $S_{z,z}\left(\omega\right)$. In the estimation
procedure, the frequency range is restricted to frequencies below
the RO peak to ensure validity of the analytical model~\eqref{eq: analytical signal S_xx}.

The optimal reconstruction of the hidden FN--PSD is shown in
Fig.~\ref{fig: method applied to simulation data}\,(c) along with
corresponding SNR estimate and the measurement noise PSD.
The PSE filter yields a significantly better reconstruction than the inverse filter method, which contains the characteristic reconstruction artifacts and deviates clearly from the hidden signal at increased measurement noise, see Fig.~\ref{fig: method applied to simulation data}\,(b).
The objective function~\eqref{eq: objective function} evaluated for the simulated stochastic data
is shown in Fig.~\ref{fig: method applied to simulation data}\,(d).
Just like in Sec.~\ref{sec: estimation method}, the objective function features a sharp minimum near at the exact value.

\section{Discussion \label{sec: Discussion}}

The method presented in Sec.~\ref{sec: estimation method} not only
provides an artifact-free reconstruction of the hidden FN--PSD, but also
allows to extract the intrinsic linewidth when it is obscured by
measurement noise. The procedure, however, relies on the specification
of the frequency-dependent SNR in the form of the analytical model
\eqref{eq: analytical signal S_xx}--\eqref{eq: analytical signal S_xixi}.
As we have demonstrated in Fig.~\ref{fig: PSE over and under estimation}, 
incorrect SNR estimates lead to reconstruction errors, which are identified
as such via inconsistencies with the assumed functional form \eqref{eq: analytical signal S_xx} of the hidden PSD.
This \emph{a priori} assumption of the functional, however, is well validated \cite{Kikuchi1989,Salvade2000,Stephan2005}, so that no false bias is imposed here.
Instead, our method exploits this additional prior knowledge about the physics of the problem to extract additional information (weak modulations of the measured PSD) from the measured data that is not used in the inverse filter method.

Even though we restricted the parameter estimation problem
in Secs.~\ref{sec: estimation method} and \ref{sec: application to simulated data} 
to a single unknown variable, it should be
straightforward to extend the method to a multivariate (nonlinear) minimization problem
where all parameters characterizing the SNR are estimated simultaneously.
Furthermore, it would be interesting to apply the estimation method in an
analogous way to the reconstruction of the RIN, which is typically more obscured by detector noise.

In principle other estimation methods can also be employed for
reconstruction of the FN--PSD from noisy time series data.
For example, Zibar \emph{et al.} \cite{Zibar2021} have used an extended
Kalman filter to estimate the effect of amplifier noise on 
the phase noise PSD of a laser.
The disadvantage of this method, however, is that it requires a
(comprehensive) mathematical model of the dynamical system under measurement,
which imposes a significant overhead.
Moreover, the application of Kalman filters to problems
with large delay (like the DSH-measurement), is notoriously difficult and computationally heavy \cite{Alexander1991, Gopalakrishnan2011}.
In contrast, the strength of parametric Wiener filters is
that they are independent of assumptions on the underlying state space model.
Moreover, since the method is formulated in the frequency-domain, it does not suffer from computational burden due to the
large delay. Finally, the method is simple to implement, as it is basically a
straightforward extension of the standard inverse filter method
(that is still contained as a limiting case).

\section{Summary}

We have presented an improved post-processing routine based on a parametric Wiener filter, that yields a potentially exact reconstruction of the FN--PSD (without any reconstruction artifacts) from DSH beat note measurements.
The method, however, requires an accurate estimate of the frequency-dependent SNR, which can be consistently obtained by deliberate suppression of the characteristic reconstruction artifacts.
In this way, both the footprint of the interferometer as well as the detector noise can be removed with high accuracy.
Remarkably, the method thus allows for the reconstruction of the intrinsic linewidth (white noise) plateau even when it is entirely obscured by measurement noise.
The approach has been demonstrated for simulated time series based on a stochastic laser rate equation model including non-Markovian $1/f$-type noise.

\appendix

\section*{Appendix}

\section{Effective Phase Measurement Noise \label{sec: phase measurement noise}} 

We seek for an approximation of the effective phase measurement noise and its two-time correlation function.
Starting from Eq.~\eqref{eq: measurement IQ}, we expand for small noise
\begin{equation*}
\arctan\left(\frac{Q\left(t\right)}{I\left(t\right)}\right)\approx\arctan\left(\tan{\left(\Phi(t)\right)}+\frac{\tan{\left(\Phi\left(t\right)\right)}}{\eta_{\mathrm{det}}\sqrt{P\left(t\right)P\left(t-\tau_{d}\right)}}\left(\frac{\xi_{Q}\left(t\right)}{\sin{\left(\Phi(t)\right)}}-\frac{\xi_{I}\left(t\right)}{\cos{\left(\Phi(t)\right)}}\right)\right)
\end{equation*}
where $\Phi\left(t\right)=\phi\left(t\right)-\phi\left(t-\tau_{d}\right)-\Delta\Omega\,t$.
Expansion to first order yields
\begin{equation*}
\arctan\left(\frac{Q\left(t\right)}{I\left(t\right)}\right)  \approx\Phi\left(t\right)+\frac{1}{\eta_{\mathrm{det}}\sqrt{P\left(t\right)P\left(t-\tau_{d}\right)}}\left(\cos{\left(\Phi\left(t\right)\right)}\xi_{Q}\left(t\right)-\sin{\left(\Phi\left(t\right)\right)}\xi_{I}\left(t\right)\right).
\end{equation*}
Expansion at the CW state with $P\left(t\right)=\overline{P}+\delta P\left(t\right)$
and $\phi\left(t\right)=\overline{\Omega}t+\delta\phi\left(t\right)$
yields
\begin{equation*}
\Delta\phi\left(t\right)=\delta\phi\left(t\right)-\delta\phi\left(t-\tau_{d}\right)\approx  \arctan\left(\frac{Q\left(t\right)}{I\left(t\right)}\right)-\overline{\Omega}\tau_{d}+\Delta\Omega\,t+\xi_{\phi}
\end{equation*}
with the effective phase measurement noise
\begin{equation*}
\xi_{\phi}\left(t\right)=\frac{1}{\eta_{\mathrm{det}}\sqrt{P\left(t\right)P\left(t-\tau_{d}\right)}}\left(\sin{\left(\Phi\left(t\right)\right)}\xi_{I}\left(t\right)-\cos{\left(\Phi\left(t\right)\right)}\xi_{Q}\left(t\right)\right).
\end{equation*}
We approximate the two-time correlation function
\begin{align*}
\left\langle \xi_{\phi}\left(t\right)\xi_{\phi}\left(t'\right)\right\rangle  & \approx\frac{1}{\eta_{\mathrm{det}}^{2}\overline{P}^{2}}\Big(\left\langle \sin{\left(\Phi\left(t\right)\right)}\sin{\left(\Phi\left(t'\right)\right)}\right\rangle \left\langle \xi_{I}\left(t\right)\xi_{I}\left(t'\right)\right\rangle \\
 & \hphantom{=\frac{1}{\eta_{\mathrm{det}}^{2}\overline{P}^{2}}\Big(}-\left\langle \cos{\left(\Phi\left(t\right)\right)}\sin{\left(\Phi\left(t'\right)\right)}\right\rangle \left\langle \xi_{Q}\left(t\right)\xi_{I}\left(t'\right)\right\rangle \\
 & \hphantom{=\frac{1}{\eta_{\mathrm{det}}^{2}\overline{P}^{2}}\Big(}-\left\langle \sin{\left(\Phi\left(t\right)\right)}\cos{\left(\Phi\left(t'\right)\right)}\right\rangle \left\langle \xi_{I}\left(t\right)\xi_{Q}\left(t'\right)\right\rangle \\
 & \hphantom{=\frac{1}{\eta_{\mathrm{det}}^{2}\overline{P}^{2}}\Big(}+\left\langle \cos{\left(\Phi\left(t\right)\right)}\cos{\left(\Phi\left(t'\right)\right)}\right\rangle \left\langle \xi_{Q}\left(t\right)\xi_{Q}\left(t'\right)\right\rangle \Big),
\end{align*}
where we have neglected photon number fluctuations and factorized
the phase and detector noise. Using $\left\langle \xi_{I}\left(t\right)\xi_{I}\left(t'\right)\right\rangle =\left\langle \xi_{Q}\left(t\right)\xi_{Q}\left(t'\right)\right\rangle =\sigma_{\mathrm{meas}}^{2}\delta\left(t-t'\right)$
and stationarity $\left\langle \xi_{I}\left(t\right)\xi_{Q}\left(t'\right)\right\rangle =\left\langle \xi_{I}\left(t'\right)\xi_{Q}\left(t\right)\right\rangle $,
we arrive at
\begin{equation*}
\left\langle \xi_{\phi}\left(t\right)\xi_{\phi}\left(t'\right)\right\rangle  \approx\frac{1}{\eta_{\mathrm{det}}^{2}\overline{P}^{2}}\left(\sigma_{\mathrm{meas}}^{2}\delta\left(t-t'\right)-\left\langle \sin{\left(\Phi\left(t\right)+\Phi\left(t'\right)\right)}\right\rangle \left\langle \xi_{I}\left(t'\right)\xi_{Q}\left(t\right)\right\rangle \right).
\end{equation*}
By neglecting the rapidly oscillating cross-correlation term, we arrive at Eq.~\eqref{eq: phase measurement noise}.

\section{Derivation of the Frequency Domain Filter Functions}

\subsection{Wiener Filter \label{sec: Wiener Filter derivation}}

We consider the mean square error between the hidden signal $x\left(t\right)$
and its reconstruction Eq.~\eqref{eq: Wiener deconvolution x estimate}
\begin{equation*}
E\left(t\right)=\left\langle \left(\hat{x}\left(t\right)-x\left(t\right)\right)^{2}\right\rangle .
\end{equation*}
Fourier transform and substitution of \eqref{eq: Wiener deconvolution - Fourier space}
yields
\begin{align*}
E\left(t\right) & =\int_{-\infty}^{\infty}\frac{\mathrm{d}\omega}{2\pi}\,\int_{-\infty}^{\infty}\frac{\mathrm{d}\omega'}{2\pi}\,\mathrm{e}^{-i\left(\omega-\omega'\right)t}\,\bigg(\left[G^{*}(\omega')H^{*}(\omega')-1\right]\left[G(\omega)H(\omega)-1\right]\left\langle X(\omega)X^{*}(\omega')\right\rangle \\
 & \phantom{=}+2\mathrm{Re}\left(G^{*}\left(\omega'\right)\left[G\left(\omega\right)H\left(\omega\right)-1\right]\left\langle X\left(\omega\right)\Xi^{*}\left(\omega'\right)\right\rangle \right)+G^{*}\left(\omega'\right)G\left(\omega\right)\left\langle \Xi\left(\omega\right)\Xi^{*}\left(\omega'\right)\right\rangle \bigg).
\end{align*}
Next, we substitute the expressions for the signal
and noise PSDs
\begin{align*}
\frac{1}{2\pi}\left\langle X\left(\omega\right)X^{*}\left(\omega'\right)\right\rangle  & =S_{x,x}\left(\omega\right)\delta\left(\omega-\omega'\right), & \frac{1}{2\pi}\left\langle \Xi\left(\omega\right)\Xi^{*}\left(\omega'\right)\right\rangle  & =S_{\xi,\xi}\left(\omega\right)\delta\left(\omega-\omega'\right),
\end{align*}
and assume uncorrelated process and measurement noise $\left\langle X \left(\omega\right) \Xi^{*}\left(\omega'\right)\right\rangle  =0$.
This yields
\begin{align*}
E\left(t\right) & =\int_{-\infty}^{\infty}\frac{\mathrm{d}\omega}{2\pi}\,\left(\left|G\left(\omega\right)H\left(\omega\right)-1\right|^{2}S_{x,x}\left(\omega\right)+\left|G\left(\omega\right)\right|^{2}S_{\xi,\xi}\left(\omega\right)\right),
\end{align*}
which is entirely independent of the time $t$. Minimization of the
reconstruction error $E\left(t\right)$ is achieved by taking the
G\^{a}teaux derivative with respect to $G\left(\omega\right)\to G\left(\omega\right)+\varepsilon\delta G\left(\omega\right)$
\begin{align*}
0\stackrel{!}{=}\lim_{\varepsilon\to0}\frac{E\left[G+\varepsilon\delta G\right]-E\left[G\right]}{\varepsilon} & =\int_{-\infty}^{\infty}\frac{\mathrm{d}\omega}{2\pi}\,\Big(\left(G\left(\omega\right)H\left(\omega\right)-1\right)H^{*}\left(\omega\right)S_{x,x}\left(\omega\right)\\
 & \hphantom{=\int_{-\infty}^{\infty}\frac{\mathrm{d}\omega}{2\pi}\,\bigg(}+G\left(\omega\right)S_{\xi,\xi}\left(\omega\right)\Big)\delta G^{*}\left(\omega\right)+\mathrm{c.c.}
\end{align*}
where the variation $\delta G\left(\omega\right)$ is arbitrary.
From this, finally, we extract the Wiener filter Eq.~\eqref{eq: standard Wiener filter}.

\subsection{Power Spectrum Equalization \label{sec: PSE Filter derivation}}

We seek for an optimal reconstruction $S_{\hat{x},\hat{x}}\left(\omega\right)$
of the PSD that minimizes the quadratic error
\begin{equation*}
E=\int_{-\infty}^{\infty}\mathrm{d}\omega\,\left(S_{\hat{x},\hat{x}}\left(\omega\right)-S_{x,x}\left(\omega\right)\right)^{2}.    
\end{equation*}
Starting from $\langle \hat{X}\left(\omega\right)\hat{X}^{*}\left(\omega'\right) \rangle = 2\pi S_{\hat{x},\hat{x}}\left(\omega\right)\delta\left(\omega-\omega'\right)$, we substitute Eq.~\eqref{eq: Wiener deconvolution - Fourier space}.
Assuming $\left\langle X\left(\omega\right)\Xi^{*}\left(\omega'\right)\right\rangle =0$, we arrive at
\begin{equation*}
S_{\hat{x},\hat{x}}\left(\omega\right)\delta\left(\omega-\omega'\right)=\left|G\left(\omega\right)\right|^{2}\left(\left|H\left(\omega\right)\right|^{2}S_{x,x}\left(\omega\right)+S_{\xi,\xi}\left(\omega\right)\right)\delta\left(\omega-\omega'\right).
\end{equation*}
The last line allows to rewrite the expression for the reconstruction
error as
\begin{equation*}
E=\int_{-\infty}^{\infty}\mathrm{d}\omega\,\left(\left(\left|G\left(\omega\right)H\left(\omega\right)\right|^{2}-1\right)S_{x,x}\left(\omega\right)+\left|G\left(\omega\right)\right|^{2}S_{\xi,\xi}\left(\omega\right)\right)^{2}.
\end{equation*}
Minimization of the error by variation of the filter $G\left(\omega\right)\to G\left(\omega\right)+\varepsilon\delta G\left(\omega\right)$
yields
\begin{align*}
0\stackrel{!}{=}\lim_{\varepsilon\to0}\frac{E\left[G+\varepsilon\delta G\right]-E\left[G\right]}{\varepsilon} & =2\int_{-\infty}^{\infty}\mathrm{d}\omega\,\left(\left|H\left(\omega\right)\right|^{2}S_{x,x}\left(\omega\right)+S_{\xi,\xi}\left(\omega\right)\right)\times\\
 & \hphantom{=2\int_{-\infty}^{\infty}}\times\left(\left(\left|G\left(\omega\right)H\left(\omega\right)\right|^{2}-1\right)S_{x,x}\left(\omega\right)+\left|G(\omega)\right|^{2}S_{\xi,\xi}(\omega)\right)\\
 & \hphantom{=2\int_{-\infty}^{\infty}}\times\left(G\left(\omega\right)\delta G^{*}\left(\omega\right)+G^{*}\left(\omega\right)\delta G\left(\omega\right)\right),
\end{align*}
from which we find Eq.~\eqref{eq: power spectrum equalization} to
be the optimal filter.

\section{Itô-Type Stochastic Differential Equations \label{sec: Ito SDEs}}

The Langevin equations \eqref{eq: Langevin laser model} can be written
as a system of Itô-type stochastic differential equations
\begin{subequations}\label{eq: Ito SDEs}
\begin{align}
\mathrm{d}P&=\left(-\gamma\left(P-P_{\mathrm{th}}\right)+\Gamma v_{g}g\left(P,N\right)P+\Gamma v_{g}g_{\mathrm{sp}}\left(P,N\right)+\sigma_{P}\left(P\right)\mathcal{F}_{P}\right)\,\mathrm{d}t\label{eq:SDEP}\\&\hphantom{=}+\sqrt{\gamma\left(1+P_{\mathrm{th}}\right)P}\,\mathrm{d}W_{\mathrm{out}}^{P}+\sqrt{\gamma P_{\mathrm{th}}\left(1+P\right)}\,\mathrm{d}W_{\mathrm{in}}^{P}+\sqrt{\Gamma v_{g}g_{\mathrm{sp}}\left(P,N\right)P}\,\mathrm{d}W_{\mathrm{st-em}}^{P}\nonumber\\
&\hphantom{=}+\sqrt{\Gamma v_{g}g_{\mathrm{abs}}\left(P,N\right)P}\,\mathrm{d}W_{\mathrm{st-abs}}^{P}+\sqrt{\Gamma v_{g}g_{\mathrm{sp}}\left(P,N\right)}\,\mathrm{d}W_{\mathrm{sp}}^{P}, \nonumber\\
\mathrm{d}\phi&=\left(\Omega_{0}+\frac{\alpha_{H}}{2}\Gamma v_{g}g\left(P,N\right)+\frac{\sigma_{P}\left(P\right)}{2P}\mathcal{F}_{\phi}\right)\,\mathrm{d}t\label{eq:SDEphi}\\&\hphantom{=}+\frac{1}{2P}\bigg(\sqrt{\gamma\left(1+P_{\mathrm{th}}\right)P}\,\mathrm{d}W_{\mathrm{out}}^{\phi}+\sqrt{\gamma P_{\mathrm{th}}\left(1+P\right)}\,\mathrm{d}W_{\mathrm{in}}^{\phi} +\sqrt{\Gamma v_{g}g_{\mathrm{sp}}\left(P,N\right)P}\,\mathrm{d}W_{\mathrm{st-em}}^{\phi}\nonumber\\
&\hphantom{=+\frac{1}{2P}\bigg(}+\sqrt{\Gamma v_{g}g_{\mathrm{abs}}\left(P,N\right)P}\,\mathrm{d}W_{\mathrm{st-abs}}^{\phi}+\sqrt{\Gamma v_{g}g_{\mathrm{sp}}\left(P,N\right)}\,\mathrm{d}W_{\mathrm{sp}}^{\phi}\bigg),\nonumber\\
\mathrm{d}N&=\left(\frac{\eta I}{q}-R\left(N\right)-\Gamma v_{g}g\left(P,N\right)P-\Gamma v_{g}g_{\mathrm{sp}}\left(P,N\right)+\sigma_{N}\left(N\right)\mathcal{F}_{N}\right)\,\mathrm{d}t\label{eq:SDEN}\\&\hphantom{=}+\sqrt{\frac{\eta I}{q}}\,\mathrm{d}W_{I}+\sqrt{R\left(N\right)}\,\mathrm{d}W_{R}-\sqrt{\Gamma v_{g}g_{\mathrm{sp}}\left(P,N\right)P}\,\mathrm{d}W_{\mathrm{st-em}}^{P} \nonumber\\
&\hphantom{=} -\sqrt{\Gamma v_{g}g_{\mathrm{abs}}\left(P,N\right)P}\,\mathrm{d}W_{\mathrm{st-abs}}^{P}-\sqrt{\Gamma v_{g}g_{\mathrm{sp}}\left(P,N\right)}\,\mathrm{d}W_{\mathrm{sp}}^{P}.\nonumber
\end{align}
\end{subequations}
Here, $\mathrm{d}W\sim\mathrm{Normal}\left(0,\mathrm{d}t\right)$
denotes the increment of the standard Wiener processes  
(Gaussian white noise) \cite{Jacobs2010}. Wiener processes with different sub- and superscripts are statistically independent.
Construction of the colored noise sources
$\mathcal{F}_{P,\phi,N}$ is described in Appendix~\ref{sec: colored noise}.

\section{Colored Noise \label{sec: colored noise}}
Colored noise sources $\mathcal{F}_{P,\phi,N}$ (subscripts are
omitted in the following) are modeled as a 
superposition of independent Ornstein--Uhlenbeck (OU) fluctuators (\emph{Markovian embedding}) \cite{Kogan1996} 
\begin{equation*}
\mathcal{F}\left(t\right)=\sqrt{\frac{A}{n}}\sum_{i=1}^{n}X_{i}\left(t\right),
\end{equation*}
where $A$ is a normalization constant (see below), $n$ is the number of OU fluctuators and
\begin{equation}
\mathrm{d}X_{i}\left(t\right)=-\gamma_{i}X_{i}\left(t\right)\,\mathrm{d}t+\sqrt{2\gamma_{i}}\,\mathrm{d}W_{i}\left(t\right). \label{eq: OU fluctuator}
\end{equation}
The fluctuators are statistically independent, \emph{i.e.}, 
$\mathrm{d}W_{i}\left(t\right)\mathrm{d}W_{j}\left(t\right)=\delta_{i,j}\,\mathrm{d}t$.
From the stationary covariance
$C_{X_{i},X_{j}}\left(\tau\right)=\left\langle X_{i}\left(t+\tau\right)X_{j}\left(t\right)\right\rangle =\delta_{i,j}\,\exp{\left(-\gamma_{i}\left|\tau\right|\right)}$, we obtain the auto-correlation function of the colored noise 
\begin{equation*}
C_{\mathcal{F},\mathcal{F}}\left(\tau\right)=\frac{A}{n}\sum_{j=1}^{n}\,\mathrm{e}^{-\gamma_{j}\left|\tau\right|}.
\end{equation*}
The corresponding PSD is obtained according to the Wiener--Khinchin
theorem \cite{Kubo1991} as
\begin{equation*}
S_{\mathcal{F},\mathcal{F}}\left(\omega\right)  =\int_{-\infty}^{\infty}\mathrm{d}\tau\,\mathrm{e}^{i\omega\tau}C_{\mathcal{F},\mathcal{F}}\left(\tau\right)=A\frac{1}{N}\sum_{j=1}^{N}\,\frac{2\gamma_{j}}{\omega^{2}+\gamma_{j}^{2}}
 =A\int_{0}^{\infty}\mathrm{d}\gamma\,\rho\left(\gamma\right)\frac{2\gamma}{\omega^{2}+\gamma^{2}},
\end{equation*}
where we introduced the continuous distribution of the relaxation rates
\begin{equation}
\rho\left(\gamma\right)=\frac{1}{n}\sum_{j=1}^{n}\delta\left(\gamma-\gamma_{j}\right). \label{eq: distribution of relaxation rates}
\end{equation}
In the following, we consider a power-law distribution
\begin{equation}
\rho\left(\gamma\right)=\frac{C_{\nu}}{\gamma^{\nu}}\Theta\left(\gamma-\gamma_{0}\right)\Theta\left(\gamma_{\infty}-\gamma\right),\qquad0<\nu<2,\label{eq: power law model}
\end{equation}
with lower and upper cutoffs $\gamma_{0}$ and $\gamma_{\infty}$.
The normalization constant $C_{\nu}=\left(1-\nu\right)/\big(\gamma_{\infty}^{1-\nu}-\gamma_{0}^{1-\nu}\big)$
ensures normalization $\int_{0}^{\infty}\mathrm{d}\gamma\,\rho\left(\gamma\right)=1$.
From Eq.~\eqref{eq: power law model}, we find
\begin{equation*}
S_{\mathcal{F},\mathcal{F}}\left(\omega\right)  =2AC_{\nu}\int_{\gamma_{0}}^{\gamma_{\infty}}\mathrm{d}\gamma\,\frac{\gamma^{1-\nu}}{\omega^{2}+\gamma^{2}}=\frac{2AC_{\nu}}{\omega^{\nu}}\int_{\gamma_{0}/\omega}^{\gamma_{\infty}/\omega}\mathrm{d}x\,\frac{x^{1-\nu}}{1+x^{2}}.
\end{equation*}
The integral can formally be solved by a hypergeometric function.
More insight, however, is gained by considering the asymptotic limit
$\gamma_{0}\to0$ and $\gamma_{\infty}\to\infty$, which
leads to
\begin{equation*}
\int_{0}^{\infty}\mathrm{d}x\,\frac{x^{1-\nu}}{1+x^{2}}=\frac{\pi}{2}\frac{1}{\sin{\left(\frac{\pi\nu}{2}\right)}}.
\end{equation*}
Hence, the PSD exhibits a power-law type frequency-dependency
\begin{equation*}
S_{\mathcal{F},\mathcal{F}}\left(\gamma_{\infty}^{-1}\ll\omega\ll\gamma_{0}^{-1}\right)  \approx\frac{1}{\omega^{\nu}},
\end{equation*}
in an arbitrarily large frequency window.
Here, we have chosen the normalization constant as
$A=\sin{\left(\frac{\pi\nu}{2}\right)}/\left(C_{\nu}\pi \right)$.
For the practical generation of time series obeying the desired PSD, it is required to approximate
the corresponding distribution
of the relaxation rates  \eqref{eq: power law model}  by finitely many $\gamma_i$. The optimal 
choice of the $n$ relaxation rates is obtained by inverse 
transform sampling.


\section*{Acknowledgments}
This work was funded by the German Research Foundation (Deutsche Forschungsgemeinschaft, DFG) under Germany's Excellence Strategy -- EXC 2046: MATH+ (Berlin Mathematics Research Center, project AA2-13).

\end{document}